\documentclass[11pt,preprint,showpacs,preprintnumbers,amsmath,amssymb]{revtex4}

\usepackage{setspace}
\usepackage{graphicx}
\usepackage{dcolumn}
\usepackage{bm}
\usepackage{multirow}
\usepackage{graphicx}
\usepackage{amssymb}
\usepackage{float}
\usepackage{CJK}
\usepackage{caption}
\usepackage{rotating}
\usepackage{subfigure}
\usepackage{relsize}
\usepackage{diagbox}
\usepackage{epstopdf}

\begin{document}

\title{Studying the $\bar{D}_1K$ molecule in the Bethe-Salpeter equation approach}

\author{Jing-Juan Qi \footnote{e-mail: qijj@mail.bnu.edu.cn}}
\affiliation{\scriptsize{Junior College, Zhejiang Wanli University, Zhejiang 315101, China}}

\author{Zhen-Yang Wang \footnote{Corresponding author, e-mail: wangzhenyang@nbu.edu.cn}}
\affiliation{\scriptsize{Physics Department, Ningbo University, Zhejiang 315211, China}}

\author{Zhu-Feng Zhang \footnote{e-mail: zhufengzhang@nbu.edu.cn}}
\affiliation{\scriptsize{Physics Department, Ningbo University, Zhejiang 315211, China}}

\author{Xin-Heng Guo \footnote{Corresponding author, e-mail: xhguo@bnu.edu.cn}}
\affiliation{\scriptsize{College of Nuclear Science and Technology, Beijing Normal University, Beijing 100875, China}}

\date{\today}

\begin{abstract}
We interpret the $X_1(2900)$ as an $S$-wave $\bar{D}_1K$ molecular state in the Bethe-Salpeter equation approach with the ladder and instantaneous approximations for the kernel. By solving the Bethe-Salpeter equation numerically with the kernel containing one-particle-exchange diagrams and introducing three different form factors (monopole, dipole, and exponential form factors) in the verties, we find the bound state exists. We also study the decay width of the decay $X_1(2900)$ to $D^-K^+$.
\end{abstract}

\pacs{**********}

\maketitle
\section{Introduction}
\label{intro}
Recently, two new open flavor states were observed by LHCb collaboration in the $D^-K^+$ invariant mass
distribution of $B^+\rightarrow D^+D^-K^+$, and the parameters are determined to be \cite{Aaij:2020ypa}
\begin{equation}
\nonumber
\begin{split}
X_0(2900):M&=2.866\pm0.007\pm0.002 \ \mathrm{GeV},\\
     \Gamma&=57\pm12\pm4 \ \mathrm{MeV},\\
X_1(2900):M&=2.904\pm0.005\pm0.001 \ \mathrm{GeV},\\
     \Gamma&=110\pm11\pm4 \ \mathrm{MeV},\\
\end{split}
\end{equation}
respectively. Since the resonances $X_0(2900)$ and $X_1(2900)$ are observed in the $D^-K^+$ channel, they should be manifestly exotic and have minimal quark content $\bar{c}du\bar{s}$. These two states are new fully open flavor states after the discovery of $X(5568)$ ($su\bar{b}\bar{d}$), which was reported by D0 collaboration in the $B_s\pi$ invariant mass distribution in 2016. However, since then LHCb, CMS, CDF, and ATLAS Collaborations have not found evidence for $X(5568)$ \cite{Aaij:2016iev,Sirunyan:2017ofq,Aaltonen:2017voc,Aaboud:2018hgx}.

In the past decades, a growing number of good candidates of exotic states have been observed, with lots of them containing $c\bar{c}$ or $b\bar{b}$ quarks \cite{Guo:2017jvc,Olsen:2017bmm}. Thus, the discovery of $X_0(2900)$ and $X_1(2900)$ have drawn a lot of attentions. The $X_0(2900)$ can be interpreted as a $cs\bar{u}\bar{d}$ compact tetraquark both in the universal quark mass picture \cite{Karliner:2020vsi} and in the quark model \cite{Wang:2020prk}, but not within an extended relativized quark model \cite{Lu:2020qmp}. In Ref. \cite{He:2020jna}, the authors used the two-body chromomagnetic interactions to find that the $X_0(2900)$ can be interpreted as a radial excited tetraquark and the $X_1(2900)$ can be an orbitally excited tetraquark. It was also suggested the $X_0(2900)$ can be interpreted as the $S$-wave $D^{\ast-}K^{\ast+}$ molecule state and the $X_1(2900)$ as the $P$-wave $\bar{c}\bar{s}ud$ compact tetraquark state \cite{Chen:2020aos}. In the chiral constituent quark model, it was shown show that no  candidate of $X(2900)$ was founded in the $IJ^P=00^+$ and $IJ^P=01^+$ $cs\bar{q}\bar{q}$ system, while there were two states in the $P$-wave excited $cs\bar{q}\bar{q}$ system, $D_1\bar{K}$ and $D_J\bar{K}$ , which could be candidates of $X(2900)$ \cite{Tan:2020cpu}. From the QCD Sum Rules, the $X_0(2900)$ and $X_1(2900)$ were studied in molecular and diquark-antidiquark tetraquark pictures, respectively, and the results for masses are in good agreement with the observed masses in the experiment \cite{Mutuk:2020igv}. Investigations bases the one-boson exchange model \cite{Liu:2020nil} and the phenomenological Lagrangian approach \cite{Huang:2020ptc}, showed that the $X_0(2900)$ can be a $D^{\ast}\bar{K}^{\ast}$ molecule, but the $X_1(2900)$ can not. In Ref. \cite{Xiao:2020ltm}, the decay width for $X_0(2900)\rightarrow\bar{D}K$ process was found to be in
agreement with the experimental data with the $S$-wave $\bar{D}^\ast K^\ast$ scenario for $X_0(2900)$ in the effective lagrangian approach. The study in Ref. \cite{Dong:2020rgs} showed that the $X_1(2900)$ as a $\bar{D}_1K$ is disfavored within the meson exchange model. In Ref. \cite{He:2020btl}, in the quasipotential Bethe-Salpeter (BS) equation approach, the authors supported the assignment of $X_0(2900)$ as a $D^{\ast}\bar{K}^{\ast}$ molecular state and $X_1(2900)$ as a $\bar{D_1}K$ virtual state.

Considering the mass of $X_1(2900)$ is about 10 MeV below the $\bar{D}_1K$ threshold, it is natural to explore the existence of the $S$-wave $\bar{D}_1K$ molecule. In this work, we will focus on the $X_1(2900)$ in the BS equation approach, investigating whether the $X_1(2900)$ can be an $S$-wave $\bar{D}_1K$ bound state. We will also to study the decay width of $X_1(2900) \rightarrow D^-K^+$.

In the rest of the manuscript we will proceed as follows. In Sec. \ref{sect-BS-PV}, we will establish the BS equation for the bound state of an axial-vector meson ($\bar{D}_1$) and a pseudoscalar meson ($K$). Then we will discuss the interaction kernel of the BS equation and calculate numerical results of the Lorentz scalar functions in the normalized BS wave function in Sec. \ref{Nor-BS}. In Sec. \ref{Decay}, the decay width of the $X_1(2900)$ to $D^-K^+$ final state will be calculated. In Sec. \ref{su},
we will present a summary of our results.

\section{The BS formalism for $\bar{D}_1K$ system}
\label{sect-BS-PV}
For the molecule composed of an axial-vector meson ($\bar{D}_1$) and a pseudoscalar meson ($K$), its BS wave function is defined as
\begin{equation}
  \chi^\mu\left(x_1,x_2,P\right) = \langle0|T\bar{D}_1^\mu(x_1)K(x_2)|P\rangle,
\end{equation}
where $\bar{D}_1(x_1)$ and $K(x_2)$ are the field operators of the axial-vector meson $\bar{D}_1$ and the pseudoscalar meson $K$ at space coordinates $x_1$ and $x_2$, respectively, $P=Mv$ is the total momentum of bound state and $v$ is its velocity. Let $m_{\bar{D}_1}$ and $m_K$ be the masses of the $\bar{D}_1$ and $K$ mesons, respectively, $p$ be the relative momentum of the two constituents, and define $\lambda_1$=$m_{\bar{D}_1}/(m_{\bar{D}_1}+m_K)$, $\lambda_2$=$m_K/(m_{\bar{D}_1}+m_K)$. The BS wave function in momentum space is defined as
\begin{equation}\label{PV-momentum-BS-function}
 \chi^\mu_P(x_1,x_2,P) = e^{-iPX}\int\frac{d^4p}{(2\pi)^4}e^{-ipx}\chi^\mu_P(p),
\end{equation}
where $ X = \lambda_1x_1 + \lambda_2x_2$ is the coordinate of the center of mass and $x = x_1 - x_2$. The momentum of the $\bar{D}_1$ meson is $p_1=\lambda_1P+p$ and that of the $K$ meson is $p_2=\lambda_2P-p$.

It can be shown that the BS wave function of the $\bar{D}_1K$ system satisfies the following BS equation:
\begin{equation}\label{PV-BS-equation}
  \chi^\mu_{P}(p)=S^{\mu\nu}_{\bar{D}_1}(p_1)\int\frac{d^4q}{(2\pi)^4}K_{\nu\lambda}(P,p,q)\chi^\lambda_{P}(q)S_{K}(p_2),
\end{equation}
where $S^{\mu\nu}_{\bar{D}_1}(p_1)$ and $S_{K}(p_2)$ are the propagators of $\bar{D}_1$ and $K$ mesons, respectively, and $K_{\nu\lambda}(P,p,q)$ is the kernel, which is defined as the sum of all the two particle irreducible diagrams with respect to $D_1$ and $K$ mesons. For convenience, in the following we use the variables $p_l (=p\cdot v)$ and $p_t(=p- p_lv)$ as the longitudinal and transverse projections of the relative momentum ($p$) along the bound state momentum ($P$), respectively. Then, in the heavy quark limit the propagator of $D_1$ is
\begin{equation}\label{D1-propagator}
  S^{\mu\nu}_{D_1}(\lambda_1P+p)=\frac{-i\left(g^{\mu\nu}-v^\mu v^\nu\right)}{2\omega_1\left(\lambda_1M+p_l-\omega_1+i\epsilon\right)},
\end{equation}
and the propagator of the $K$ meson is
\begin{equation}\label{K-propagator}
  S_K(\lambda_2P-p)=\frac{i}{\left(\lambda_2M-p_l\right)^2-\omega_2^2+i\epsilon},
\end{equation}
respectively, where $\omega_{1(2)} = \sqrt{m_{\bar{D}_1(K)}^2+p_t^2}$ (we have defined $p_t^2=-p_t\cdot p_t$).

In the BS equation approach, the interaction between $\bar{D}_1$ and $K$ mesons arises from the light vector-meson ($\rho$ and $\omega$) exchange. Based on the heavy quark symmetry and the chiral symmetry, the relevant effective Lagrangian used in this work is shown in the following \cite{Ding:2008gr}:
\begin{equation}
\begin{split}\label{D1K-Lagrangian}
  \mathcal{L}_{D_1D_1V} =&ig_{D_1D_1V}(D^\nu_{1b}\overleftrightarrow{\partial}_\mu D_{1a\nu}^\dag)V_{ba}^\mu+ig'_{D_1D_1V}(D_{1b}^\mu D_{1a}^{\nu\dag}-D_{1a}^{\mu\dag}D_{1a}^\nu)(\partial_\mu V_\nu-\partial_\nu V_\mu)_{ba}\\
  &+ig_{\bar{D}_1\bar{D}_1V}(\bar{D}_{1b\nu}\overleftrightarrow{\partial}_\mu \bar{D}_{1a}^{\nu\dag})V_{ab}^\mu+ig'_{\bar{D}_1\bar{D}_1V}(\bar{D}_{1b}^\mu \bar{D}_{1a}^{\nu\dag}-\bar{D}_{1a}^{\mu\dag}\bar{D}_{1b}^\nu)(\partial_\mu V_\nu-\partial_\nu V_\mu)_{ab}\\
  \mathcal{L}_{KKV} =&ig_{KKV}(K_b\overleftrightarrow{\partial}_\mu K_a^\dag)V_{ba}^\mu+ig_{\bar{K}\bar{K}V}(\bar{K}_b\overleftrightarrow{\partial}_\mu\bar{K}_a^\dag)V_{ba}^\mu,\\
  \end{split}
\end{equation}
where $a$ and $b$ represent the light flavor quark ($u$ and $d$), $V_\mu$ is a $3\times3$ Hermitian matrix containing $\rho$, $\omega$, $K^\ast$, and $\phi$:
\begin{eqnarray}
V&=&\left(\begin{array}{ccc}
\frac{\rho^{0}}{\sqrt{2}}+\frac{\omega}{\sqrt{2}}&\rho^{+}&K^{*+}\\
\rho^{-}&-\frac{\rho^{0}}{\sqrt{2}}+\frac{\omega}{\sqrt{2}}&
K^{*0}\\
K^{*-} &\bar{K}^{*0}&\phi
\end{array}\right).\label{vector}
\end{eqnarray}
The coupling constants involved in Eq. (\ref{D1K-Lagrangian}) are related to each other as follows \cite{Ding:2008gr}:
\begin{equation}
\begin{split}
&g_{D_1D_1V}=-g_{\bar{D}_1\bar{D}_1V}=\frac{1}{\sqrt{2}}\beta_2 g_V,\\
&g'_{D_1D_1V}=-g'_{\bar{D}_1\bar{D}_1V}=\frac{5\lambda_2g_V}{3\sqrt{2}}m_{D_1},\\
&g_{KKV}=g_V/2,\\
\end{split}
\end{equation}
where the parameters $\beta_2g_V$ and $\lambda_2g_V$ are given by $2g_{\rho NN}$ and $\frac{3}{10m_N}(g_{\rho NN}+f_{\rho NN})$, respectively, with $g_{\rho NN}^2/4\pi=0.84$ and $f_{\rho NN}/g_{\rho NN}=6.10$ \cite{Wang:2019aoc}. The parameter $g_V=5.8$ is determined by the Kawarabayashi-Suzuki-Riazuddin-Fayyazuddin relations \cite{Ding:2008gr}.

\begin{figure}[ht]
\centering
    \rotatebox{0}{\includegraphics*[width=0.4\textwidth]{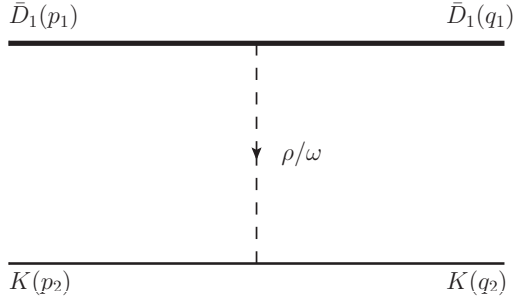}}
    \caption{One-particle exchange diagrams induced by vector mesons $\rho$ and $\omega$.}
  \label{bound-state}
\end{figure}

Then, at the tree level, in the $t$-channel the kernel for the BS equation of the $\bar{D}_1K$ system in the lader approximation includes the following term (see Fig. \ref{bound-state}):
\begin{equation}\label{D1D-kernel}
\begin{split}
  K^{\tau\sigma}_{direct}(P,p,q;m_V)=&-(2\pi)^4\delta^4(p'_1+p'_2-p_1-p_2)c_I \Big\{g_{D_1D_1V}g_{DDV}(p_1+q_1)_\gamma(p_2+q_2)_\rho g^{\tau\sigma}\\
  &\times\Delta^{\rho\gamma}(k,m_V)+g'_{D_1D_1V}g_{DDV}(p_2+q_2)_\rho\left[k^\tau\Delta^{\rho\sigma}(k,m_V)-k^\sigma\Delta^{\rho\tau}(k,m_V)\right]\Big\},\\
\end{split}
\end{equation}
where $m_V$ ($V=\rho, \omega$) represents the mass of the exchanged light vector meson $\rho$ or $\omega$ , $c_I$ is the isospin coefficient: $c_0 =3, 1$ and $c_1 =-1, 1$ for $\rho$ and $\omega$, respectively, $\Delta^{\mu\nu}$ represents the propagator for the light vector meson.

In order to describe the phenomena in the real world, we should include a form factor at each interacting vertex of hadrons to include the finite-size effects of these hadrons. For the meson-exchange case, the form factor is assumed to take the following forms:
\begin{equation}\label{form-factor}
\begin{split}
  F_M(k)&=\frac{\Lambda_M^2-m^2}{\Lambda_M^2-k^2}, \\
  F_D(k)&=\frac{(\Lambda_D^2-m^2)^2}{(\Lambda_D^2-k^2)^2}, \\
  F_E(k)&=e^{(k^2-m^2)/\Lambda_E^2}, \\
\end{split}
\end{equation}
in the monopole ($M$), dipole ($D$), and exponential ($E$) models, respectively, where $\Lambda$, $m$ and $k$ represent the cutoff parameter, the mass of the exchanged meson and the momentum of the exchanged meson, respectively. The value of $\Lambda$ is near 1 GeV which is the typical chiral symmetry breaking scale.

In general, for an axial-vector meson ($D_1$) and a pseudoscalar meson ($K$) bound state, the BS wave function $\chi_P^\mu(p)$ has the following form:
\begin{equation}
\chi_P^\mu(p)=f_0(p)p^\mu+f_1(p)P^\mu+f_2(p)\epsilon^\mu+f_3(p)\varepsilon^{\mu\nu\alpha\beta}p_\alpha P_\beta\epsilon_\nu,
\end{equation}
where $f_i(p)$ $(i = 0,1,2,3)$ are Lorentz-scalar functions and $\epsilon^\mu$ represents the polarization vector of the bound state. After considering the constraints imposed by parity and Lorentz transformations, it is easy to prove that $\chi_P^\mu(p)$ can be simplified as
\begin{equation}\label{BS-wave-function}
  \chi_P^\mu(p)=f(p)\varepsilon^{\mu\nu\alpha\beta}p_\alpha P_\beta\epsilon_\nu,
\end{equation}
where the scalar function $f(p)$ contains all the dynamics.

In the following derivation of the BS equation, we will apply the instantaneous approximation, in which the energy exchanged between the constituent particles of the binding system is neglected. In our calculation we choose the absolute value of the binding energy $E_b$ of the $\bar{D}_1K$ system (which is defined as $E_b=M-m_{D_1}-m_K$) less than 30 MeV. In this case the exchange of energy between the constituent particles can be neglected.

Substituting Eqs. (\ref{D1-propagator}), (\ref{K-propagator}), (\ref{D1D-kernel}) and (\ref{form-factor}) into Eq. (\ref{PV-BS-equation}) and using the covariant instantaneous approximation in the kernel, $p_l=q_l$, one obtains the folowing expression:
\begin{equation}\label{4-p-BS-equation}
  \begin{split}
  f(p)=&\int\frac{d^4q}{(2\pi)^4}\frac{i}{6\omega_1(\lambda_1M+p_l-\omega_1+i\epsilon)[(\lambda_2M-p_l)^2-\omega_2^2+i\epsilon][-(p_t-q_t)^2-m_V^2]}\\
       &\Big\{g_{\bar{D}_1\bar{D}_1V}g_{KKV}\left[4(\lambda_1M+p_l)(\lambda_2M-p_l)+(p_t+q_t)^2+(p_t^2-q_t^2)^2/m_V^2\right]\\
       &+g'_{\bar{D}_1\bar{D}_1V}g_{KKV}\omega_2(p_t\cdot q_t-q_t^2)/(\lambda_2M-\omega_2))\Big\}F^2(k_t)f(q),
  \end{split}
\end{equation}
where $k_t=p_t-q_t$ is the momentum of the exchanged meson in the covariant instantaneous approximation.

In Eq. (\ref{4-p-BS-equation}) there are poles in the plane of $p_l$ at $-\lambda_1 M+\omega_1-i\epsilon$, $\lambda_2 M+\omega_2-i\epsilon$ and $\lambda_2 M-\omega_2+i\epsilon$. By choosing the appropriate contour, we integrate over $p_l$ on both sides of Eq. (\ref{4-p-BS-equation}) in the rest frame of the bound state, then we obtain the following equation:
\begin{equation}\label{3-dimension-BS-equation}
  \begin{split}
   \tilde{f}(p_t) =&\int\frac{dq_t^3}{(2\pi)^3}\frac1{12\omega_1\omega_2(M-\omega_1-\omega_2)\left[-(p_t-q_t)^2-m_V^2\right]}\\
   &\times\Big\{3g_{\bar{D}_1\bar{D}_1V}g_{KKV}\left[4\omega_2(M-\omega_2)+(p_t+q_t)^2+(p_t^2-q_t^2)^2/m_V^2\right]\\
   &+2g'_{\bar{D}_1\bar{D}_1V}g_{KKV}\omega_2(p_t\cdot q_t-q_t^2)/(\lambda_2M-\omega_2)\Big\}F^2(k_t)\tilde{f}(q_t),
  \end{split}
\end{equation}
where $\tilde{f}(p_t)\equiv\int dp_l f(p)$.

Now, we can solve the BS equation numerically and study whether the $S$-wave $\bar{D}_1K$ bound state exists or not. It can be seen from Eq. (\ref{3-dimension-BS-equation}) that there is only one free parameter in our model, the cutoff $\Lambda$, which enters through various phenomenological form factors in Eq. (\ref{form-factor}).  It contains the information about the extended interaction due to the structures of hadrons. The value of $\Lambda$ is of order 1 GeV which is the typical scale of nonperturbative QCD interaction. In this work, we shall treat $\Lambda$ as a parameter and vary it in a wide range 0.8-4.8 GeV when the binding energy $E_b$ is in the region from -5 to -30 MeV to see if the BS equation has solutions.

To find out the possible molecular bound states, one only needs to solve the homogeneous BS equation.  One numerical solution of the homogeneous BS equation corresponds to a possible bound state.  The integration region in each integral is discretized into $n$ pieces, with $n$ being sufficiently large. In this way, the integral equation is converted into an $n\times n$ nmatrix equation, and the scalar wave function will now be regarded as an $n$-dimensional vector. Then, the integral equation can be illustrated as $\tilde{f}^{(n)}(p_t)=A^{(n\times n)}(p_t,q_t)\tilde{f}^{(n)}(q_t)$, where $\tilde{f}^{(n)}(p_t) (\tilde{f}^{(n)}(q_t))$ is an $n$-dimensional vector, and $A^{(n\times n)}(p_t,q_t)$ is an $n \times n$ matrix, which corresponds to the matrix labeled by $p_t$ and $q_t$ in each integral equation. Generally, $p_t$ ($q_t$) varies from 0 to $+\infty$. Here, $p_t$ ($q_t$) is transformed into a new variable $t$ that varies from $-1$ to 1 based on the Gaussian integration method,
\begin{equation}
p_t=\mu+w\log\left[1+y\frac{1+t}{1-t}\right],
\end{equation}
where $\mu$ is a parameter introduced to avoid divergence in numerical calculations, $w$ and $y$ are parameters used in controlling the slope of wave functions and finding the proper solutions
for these functions. Then one can obtain the numerical results of the BS wave functions by requiring the eigenvalue of the eigenvalue equation to be 1.0.

In our calculation, we choose to work in the rest frame of the bound state in which $P=(M,0)$. We take the averaged masses of the mesons from the PDG \cite{Zyla:2020zbs}, $m_{D_1}=2420.8$ MeV, $m_{K}=494.98$ MeV, $m_\rho=775.26$ MeV, and $m_\omega=782.65$ MeV. After searching for possible solutions in the isoscalar channel of the $\bar{D}_1K$ system, we find the bound state exists. We list some values of $E_b$ and the corresponding $\Lambda$ for the three different form factor models in Table \ref{Eb-Lambda}.

\begin{table}[tb]
\renewcommand{\arraystretch}{1.2}
\centering
\caption{
Values of $E_b$ and corresponding cutoff $\Lambda_M$, $\Lambda_D$, and $\Lambda_E$ for $I=0$ and $I=1$ $\bar{D}_1K$ bound states for the monopole, dipole, and exponential form factors, respectively.}
\begin{tabular*}{\textwidth}{@{\extracolsep{\fill}}cccccccc}
\hline
\hline
                    &$E_b$(MeV)        &  -5    &  -10   &  -15   &  -20   &  -25   &  -30  \\
\hline
\multirow{3}{*}{I=0}&$\Lambda_M$(MeV)  &  1208  &  1261  &  1297  &  1327  &  1352  &  1375   \\
                    &$\Lambda_D$(MeV)  &  1668  &  1756  &  1817  &  1867  &  1910  &  1948  \\
                    &$\Lambda_E$(MeV)  &  1159  &  1231  &  1280  &  1321  &  1356  &  1386  \\
\multirow{3}{*}{I=1}&$\Lambda_M$(MeV)  &  1541  &  1649  &  1723  &  1783  &  1835  &  1880   \\
                    &$\Lambda_D$(MeV)  &  1897  &  2371  &  2492  &  2589  &  2671  &  2744  \\
                    &$\Lambda_E$(MeV)  &  1571  &  1708  &  1804  &  1881  &  1947  &  2006  \\
\hline
\hline
\end{tabular*}\label{Eb-Lambda}
\end{table}

\section{The Normalization Condition of the BS wave function}
\label{Nor-BS}
To find out whether the bound state of the $\bar{D}_1K$ system exists or not, one only needs to solve the homogeneous BS equation. However, when we want to calculate physical quantities such as the decay width we have to face the problem of the normalization of the BS wave function. In the following we will discuss the normalization of the BS wave function $\chi^\mu_P(p)$.

In the heavy quark limit, the normalization of the BS wave function of the $\bar{D}_1K$ system can be written as \cite{Guo:2007qu}
\begin{equation}
i\int\frac{d^4pd^4q}{(2\pi)^8}\bar{\chi}^\mu_P(p)\frac{\partial}{\partial P_0}[I_{P\mu\nu}(p,q)]\chi^\nu(q)=2E_P,
\end{equation}
where $I_{P\mu\nu}(p,q)=(2\pi)^4\delta^4(p-q)S^{-1}_{\mu\nu}(p_1)S^{-1}(p_2)$.

In the rest frame, the normalization condition can be written in the following form:
\begin{equation}
\begin{split}
-i\int\frac{d^4p}{(2\pi)^4}\Big\{&4M^2p_t^2\big[\lambda_1^2(6\lambda_2^2M^2-6\lambda_2Mp_l+p_l^2-\omega_2^2)\\
&+2\lambda_1\lambda_2p_l(3\lambda_2M-2p_l)+\lambda_2^2(p_l^2-\omega_1^2)\big]f^2(q)=1.
\end{split}
\end{equation}
From Eqs. (\ref{4-p-BS-equation}) and (\ref{3-dimension-BS-equation}), we obtain
\begin{equation}
  \begin{split}
  f(p)=&\frac{i\omega_2(M-\omega_1-\omega_2)}{\pi(\lambda_1M+p_l-\omega_1+i\epsilon)(\lambda_2M-p_l+\omega_2-i\epsilon)(\lambda_2M-p_l-\omega_2+i\epsilon)}\tilde{f}(p_t).\\
  \end{split}
\end{equation}
Then, one can recast the normalization condition for the BS wave function into the form
\begin{equation}\label{3-Nor-Eq}
\begin{split}
&-\int\frac{d^3p_t}{8\pi^5}\frac{M^2p_t^2\omega_1}{\omega_2^2(M-\omega_1-\omega_2)^2}\Big\{\lambda_2^2(p_t^2-\omega_1^2)(\lambda_2M-\omega_1-3\omega_2)+\lambda_1^3(\lambda_2^2M^3-2M\omega_2^2)\\
&+\lambda_1\lambda_2[2\lambda_2^3M^3+\lambda_2M(p_t^2-\omega_1^2)-4\omega_2^2(\omega_1-\omega_2)-2\lambda_2^2M^2(\omega_1+3\omega_2)]\\
&+\lambda_1^2[3\lambda_2^3M^3-6\lambda_2M\omega_2^2+2\omega_2^2(\omega_1+\omega_2)-\lambda_2^2M^2(\omega_1+3\omega_2)]\Big\}\tilde{f}^2(p_t)=1.
\end{split}
\end{equation}

The wave function obtained in the previous section (which is calculated numerically from Eq.(\ref{3-dimension-BS-equation})) can be normalized by Eq. (\ref{3-Nor-Eq}).

\begin{figure}[htbp]
\centering
\subfigure[]{
\begin{minipage}[t]{0.3\linewidth}
\centering
\includegraphics[width=2.1in]{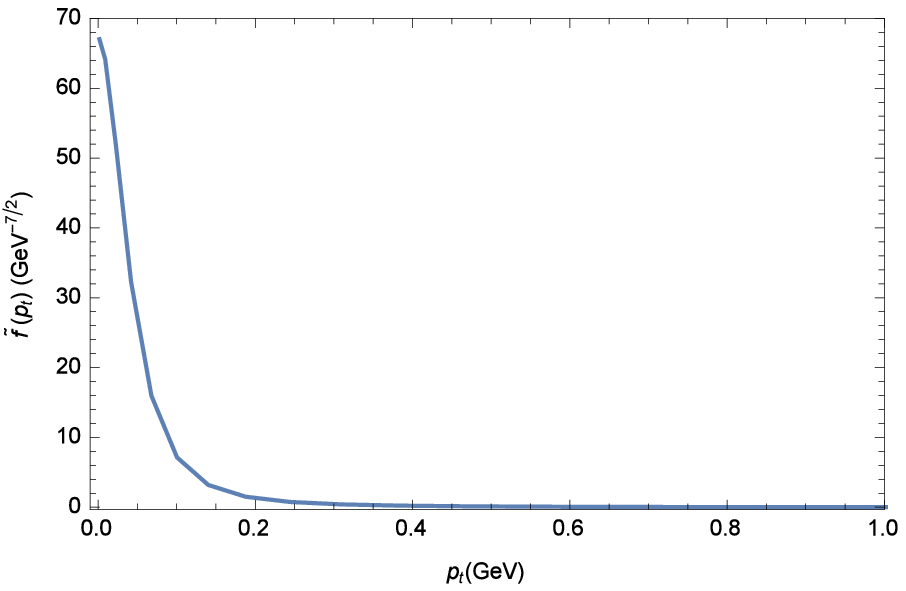}
\end{minipage}%
}%
\subfigure[]{
\begin{minipage}[t]{0.3\linewidth}
\centering
\includegraphics[width=2.1in]{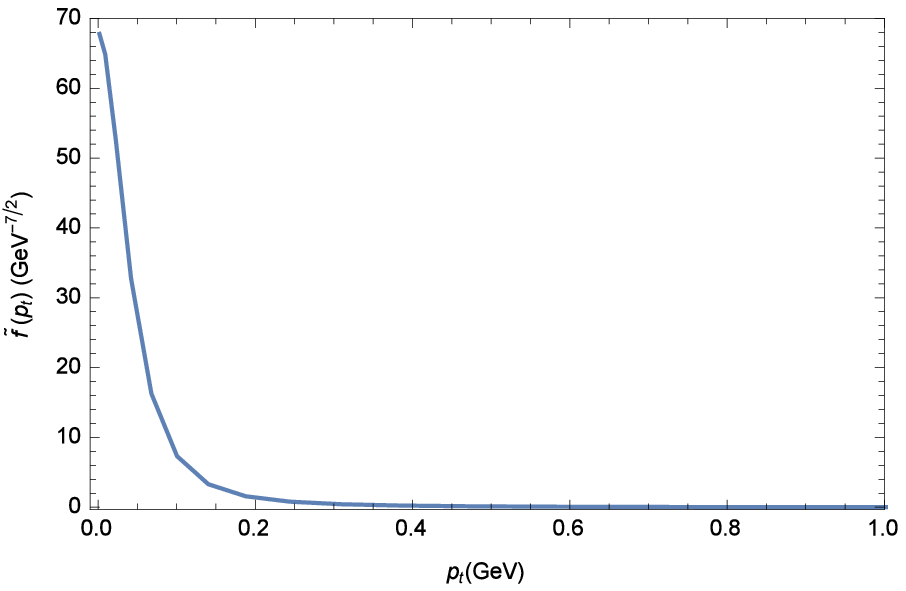}
\end{minipage}
}%
\subfigure[]{
\begin{minipage}[t]{0.3\linewidth}
\centering
\includegraphics[width=2.1in]{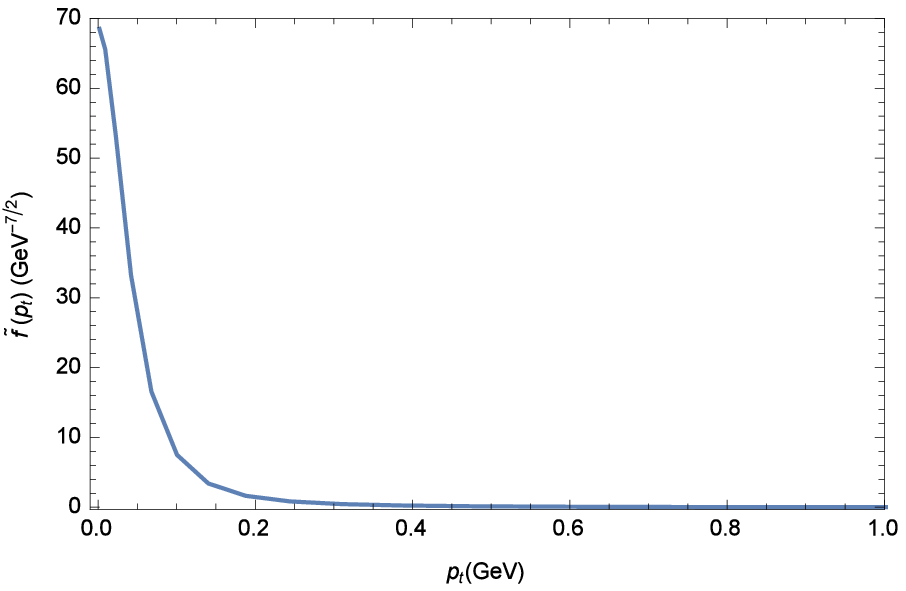}
\end{minipage}%
}%
\centering
\caption{Numerical results of the normalized Lorentz scalar function $\tilde{f}(p_t)$ for the $X_{1}(2900)$ in the $I=0$ $\bar{D}_1K$ molecular picture with (a) the monopole form factor, (b) the dipole form factor, and (c) the exponential form factor.}
\label{W0-bound-state}
\end{figure}

\begin{figure}[htbp]
\centering
\subfigure[]{
\begin{minipage}[t]{0.3\linewidth}
\centering
\includegraphics[width=2.1in]{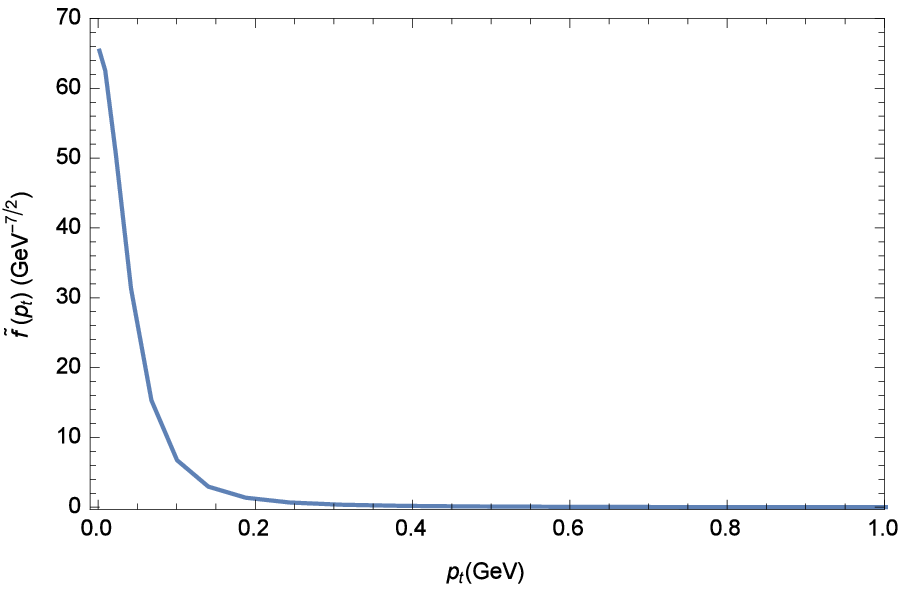}
\end{minipage}%
}%
\subfigure[]{
\begin{minipage}[t]{0.3\linewidth}
\centering
\includegraphics[width=2.1in]{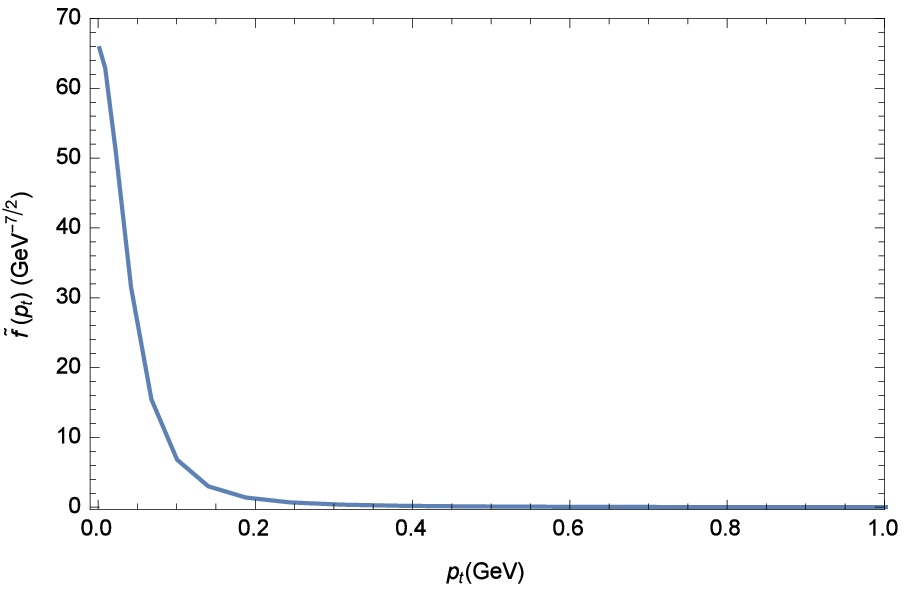}
\end{minipage}
}%
\subfigure[]{
\begin{minipage}[t]{0.3\linewidth}
\centering
\includegraphics[width=2.1in]{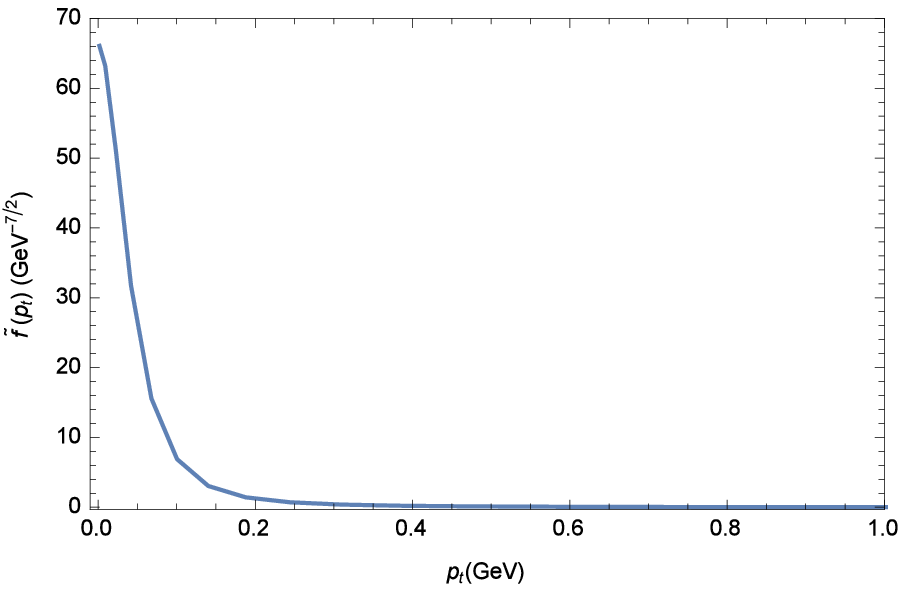}
\end{minipage}%
}%
\centering
\caption{Numerical results of the normalized Lorentz scalar function $\tilde{f}(p_t)$ for the $X_{1}(2900)$ in the $I=1$ $\bar{D}_1K$ molecular picture with (a) the monopole form factor, (b) the dipole form factor, and (c) the exponential form factor.}
\label{W1-bound-state}
\end{figure}

In our case, the binding energy $E_b=M_{X_{1}(2900)}-(M_{{\bar{D}_1}}+M_K)\simeq -12.4$ MeV, where we have used the mass of the $X_{1}(2900)$ as 2904 MeV. From our calculations, we find the $I=0$ $\bar{D}_1K$ system can be assigned as the $X_{1}(2900)$ state when the cutoff $\Lambda$ = 1280 MeV, 1788 MeV, and 1257 MeV for the monopole, dipole, and exponential form factors, respectively, the $I=1$ $\bar{D}_1K$ system can be the $X_{1}(2900)$ state when the cutoff $\Lambda$ = 1688 MeV, 2434 MeV, and 1758 MeV for the monopole, dipole, and exponential form factors, respectively. The corresponding numerical results of the normalized Lorentz scalar function, $\tilde{f}(p_t)$, are given in Figs. \ref{W0-bound-state} and \ref{W1-bound-state} for the $X_{1}(2900)$ states with $I=0$ and $I=1$ in the $\bar{D}_1K$ molecule picture for the monopole, dipole, and exponential form factors, respectively.

\section{The decay of $X_{1}(2900)\rightarrow D^-K^+$}
\label{Decay}

\begin{figure}[ht]
\centering
    \rotatebox{0}{\includegraphics*[width=0.4\textwidth]{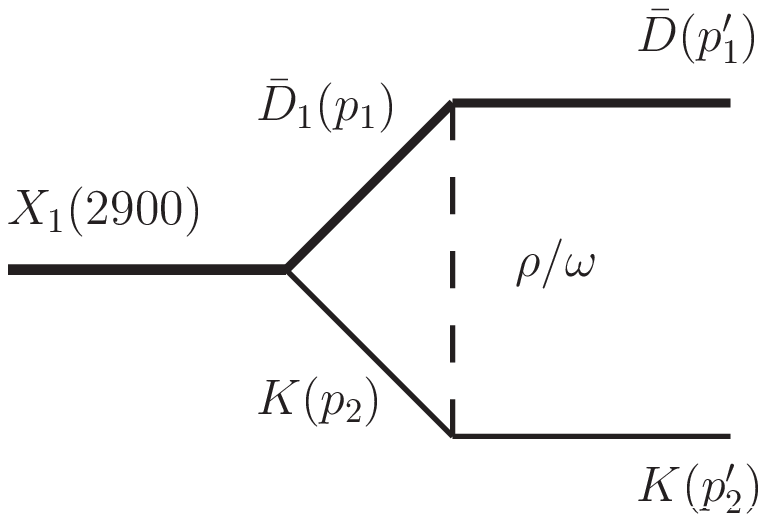}}
    \caption{The diagrams contributing to the $X_{1}(2900)\rightarrow D^-K^+$ decay process induced by $\rho$ and $\omega$.}
  \label{decay}
\end{figure}

Besides investigating whether the bound state of the $\bar{D}_1K$ system can be $X_1(2900)$ or not, we can also study the decay of the $X_1(2900)$ as the $S$-wave $\bar{D}_1K$ bound state. The $X_1(2900)$ can decay to $D^-K^+$ via the Feynman diagrams in Fig. \ref{decay}. which are induced by the effective Lagrangians for the $D_1DV$ and $KKV$ vertices (which have been given in Eq. (\ref{D1K-Lagrangian})) as the following \cite{Ding:2008gr}:
\begin{equation}
\begin{split}
\mathcal{L}_{DD_1V}=&g_{DD_1V}D_{1b}^\mu V_{\mu ba}D_a^\dag+g'_{DD_1V}(D_{1b}^\mu\overleftrightarrow{\partial}^\nu D_a^\dag)(\partial_\mu V_\nu-\partial_\nu V_\mu)_{ba}\\
  &+g_{\bar{D}\bar{D}_1V}\bar{D}_a^\dag V_{\mu ba}\bar{D}_{1b}^\mu+g'_{\bar{D}\bar{D}_1V}(\bar{D}_{1b}^\mu\overleftrightarrow{\partial}^\nu \bar{D}_a^\dag)(\partial_\mu V_\nu-\partial_\nu V_\mu)_{ba}+H.c.,\\
\end{split}
\end{equation}
where the coupling constants are given as $g_{DD_1V}=-g_{\bar{D}\bar{D}_1V}=-\frac2{\sqrt{3}}\zeta_1g_v\sqrt{m_Dm_{D_1}}$, $g'_{DD_1V}=-g'_{\bar{D}\bar{D_1}V}=\frac1{\sqrt{3}}\mu_1g_V$ \cite{Ding:2008gr}, with the two parameters $\zeta_1$ and $\mu_1$ being involved in the coupling constants, about which the information is very scarce leading them undetermined. However, in the heavy quark limit, we can roughly assume that the coupling constants $g_{D{D_1}V}$ and $g'_{DD_1V}$ are equal to $g_{D^\ast D_0V}$ (=$\zeta g_V\sqrt{2m_{D^\ast}m_{D^0}}$) and $g'_{D^\ast D_0V}$ (=$1/\sqrt{2}\mu g_V$), respectively. The parameters $\mu=0.1$ $\mathrm{GeV}^{-1}$ and $\zeta=0.1$ are taken in Ref. \cite{Casalbuoni:1996pg}.

According to the above interactions, we can write down the amplitude for the decay $X_1(2900)\rightarrow D^-K^+$ induced by light vector meson ($\rho$ and $\omega$) exchanges as shown in Fig. \ref{decay}, as the following:
\begin{equation}
\begin{split}
\mathcal{M}=&g_{KKV}\sqrt{2E}\int\frac{d^4p}{(2\pi)^4}F^2(k)\big[g_{\bar{D}\bar{D}_1V}(p_2+p'_2)^\alpha\Delta_{\alpha\mu}(k,m_V)\\
&+g'_{\bar{D}\bar{D}_1V}(p_1+p'_1)^\nu(p_2+p'_2)^\alpha\left(k_\mu\Delta_{\alpha\nu}(k,m_V)-k_\nu\Delta_{\alpha\mu}(k,m_V)\right)\big]\chi_{P}^\mu(p)
\end{split}
\end{equation}

In the rest frame, we define $p_1'=(E_1',-\mathbf{p}'_1)$ and $p_2'=(E_2',\mathbf{p}'_2)$ to be the momenta of $D$ and $K$, respectively. According to the kinematics of the two-body decay of the initial state in the rest frame, one has
\begin{equation}
\begin{split}
E_1'&=\frac{M^2-m_2^{'2}+m_1^{'2}}{2M},\quad\quad E_1'=\frac{M^2-m_1^{'2}+m_2^{'2}}{2M},\\
|\mathbf{p}'_1|&=|\mathbf{p}'_2|=\frac{\sqrt{[M^2-(m'_1+m'_2)^2][M^2-(m'_1-m'_2)^2]}}{2M},
\end{split}
\end{equation}
and
\begin{equation}
d\Gamma=\frac{1}{32\pi^2}|\mathcal{M}|^2\frac{|\mathbf{p}'|}{M^2}d\Omega,
\end{equation}
where $|\mathbf{p}'_1|$ and $|\mathbf{p}'_2|$ are the norm of the 3-momentum of the particles in the final states in the rest frame of the initial bound state and $\mathcal{M}$ is the Lorentz-invariant decay amplitude of the process.

Substituting the normalized numerical solutions of the BS equation, and the cutoff $\Lambda$ are 1280 MeV, 1788 MeV, and 1257 MeV with $I=0$ and 1688 MeV, 2434 MeV, and 1758 MeV with $I=1$ for the monopole, dipole, and exponential form factors,
respectively. The decay widths of the $X_1(2900)$ to $D^-K^+$  can be obtained as following:
\begin{equation}
\Gamma_{X_1(2900)(I=0)\rightarrow D^-K^+}=\left\{
\begin{aligned}
 & 70.73\ \mathrm{MeV}\ \mathrm{with\ monopole\ form\ factor}, \\
 & 98.75\ \mathrm{MeV}\ \mathrm{with\ dipole\ form\ factor}, \\
 & 60.38\ \mathrm{MeV}\ \mathrm{with\  exponential\ form\ factor},
\end{aligned}
\right.
\end{equation}
and
\begin{equation}
\Gamma_{X_1(2900)(I=1)\rightarrow D^-K^+}=\left\{
\begin{aligned}
 & 28.14\ \mathrm{MeV}\ \mathrm{with\ monopole\ form\ factor}, \\
 & 18.13\ \mathrm{MeV}\ \mathrm{with\ dipole\ form\ factor}, \\
 & 12.78\ \mathrm{MeV}\ \mathrm{with\  exponential\ form\ factor}.
\end{aligned}
\right.
\end{equation}

From our calculation results, we can see that different form factors have a great influence on the decay width, and different cutoff $\Lambda$ for the same form factor also have a great influence on the decay width.

\section{summary}
\label{su}
In this paper, we studied the $X_1(2900)$ with the hadronic molecule interpretation by regarding it as a bound state of $\bar{D}_1$ and $K$ mesons in the BS equation approach. In our model, we applied the ladder and instantaneous approximations to obtain the kernel containing one-particle-exchange diagrams and introduced three different form factors (the monopole form factor, the dipole form factor, and the exponential form factor) at the interaction vertices. From the calculating results we find that there exist bound states of the $\bar{D}_1K$ system. The binding energy depends on the value of the cutoff $\Lambda$. For the $I=0$ $\bar{D}_1K$ system, we find the cutoff regions in which the solutions (with the binding energy $E_b\in$ (-5, -30) MeV) for the ground state of the BS equation can be found (in units of MeV): $\Lambda_M\sim$ (1208, 1375), $\Lambda_D\sim$ (1668, 1948), and $\Lambda_E\sim$ (1159, 1386) for the monopole form factor, the dipole form factor, and the exponential form factor, respectively. For the $I=1$ $\bar{D}_1K$ system, we find two the regions (in units of MeV): $\Lambda_M\sim$ (1541, 1880), $\Lambda_D\sim$ (1897, 2744), and $\Lambda_E\sim$ (1571, 2006) in which the solutions of the BS equation can be found. Thus, we can confirm that $X_1(2900)$ can be regarded as the $S$-wave $\bar{D}_1K$ molecules in our model.

We applied the numerical solutions for the BS wave functions with the corresponding cutoff ($\Lambda$ = 1280 MeV, 1788 MeV, and 1257 MeV for $I=0$ and $\Lambda$ = 1688 MeV, 2434 MeV, and 1758 MeV for $I=1$ with the monopole, dipole,
and exponential form factors, respectively.) to calculate the decay widths of $X_{1}(2900)\rightarrow D^-K^+$ induced by $\rho$ and $\omega$ exchanges. We predict the decay widths are 70.73, 98.75, and 60.38 MeV and 28.14, 18.13, and 12.78 MeV for $X_1(2900)$ as $I=0$ and $I=1$ $\bar{D}_1K$ molecules with the corresponding cutoff in the decay process, respectively. From our study, the $X_1(2900)$ is suitable as $I=0$ $\bar{D}_1K$ molecular state. There are two uncertain factors in the calculation of the decay width, one is that the parameters $\xi_1$ and $\mu_1$ have not been determined since the information about them is very scarce, the other is that we can not give the definite value of the cutoff $\Lambda$.

\acknowledgments
This work was supported by National Natural Science Foundation of China (Projects No. 11775024, No.11605150 and No.11947001), the Natural Science Foundation of Zhejiang pvovince (No. LQ21A050005), the Ningbo Natural Science Foundation (No.2019A610067), the Fundamental Research Funds for the Provincial
Universities of Zhejiang Province and K.C.Wong Magna Fund in Ningbo University.


\end{document}